\documentstyle[preprint,aps,epsfig]{revtex}
\newcommand{\lsim}{\mathrel{\rlap{\lower4pt\hbox{\hskip0pt$\sim$}}
\raise1pt\hbox{$<$}}}
\newcommand{\gsim}{\mathrel{\rlap{\lower4pt\hbox{\hskip0pt$\sim$}}
\raise1pt\hbox{$>$}}}
\tighten
\begin{document}

\title{Mean field exponents and small quark masses}

\author{A.~H\"oll\footnotemark[2], 
P. Maris\footnotemark[3] and 
C.D. Roberts\footnotemark[3]\vspace*{0.2\baselineskip}}
\address{\footnotemark[2]Fachbereich Physik, Universit\"at Rostock,
D--18051 Rostock, Germany\\\vspace*{0.2\baselineskip}
\footnotemark[3]Physics Division, Building 203, Argonne National
Laboratory, Argonne Illinois 60439-4843, USA }
\date{Pacs Numbers: 11.10.Wx, 12.38.Mh, 24.85.+p, 05.70.Fh }

\maketitle

\begin{abstract}
We demonstrate that the restoration of chiral symmetry at finite-$T$ in a
class of confining Dyson-Schwinger equation (DSE) models of QCD is a mean
field transition, and that an accurate determination of the critical
exponents using the chiral and thermal susceptibilities requires very small
values of the current-quark mass: $\log_{10}(m/m_u)\lsim -5$.  Other classes
of DSE models characterised by qualitatively different interactions also
exhibit a mean field transition.  Incipient in this observation is the
suggestion that mean field exponents are a result of the gap equation's
fermion substructure and not of the interaction.
\end{abstract}
It is anticipated that the restoration of chiral symmetry, which accompanies
the formation of a quark-gluon plasma at finite temperature, $T$, is an
equilibrium, second-order phase transition.  Such transitions are completely
characterised by two critical exponents: $(\beta,\delta)$,
%
%
which describe the response of any one of the equivalent chiral order
parameters, ${\cal X}$, to changes in $T$ and in the current-quark mass, $m$.
Denoting the critical temperature by $T_c$, and introducing the
reduced-temperature $t:= T/T_c-1$ and reduced mass $h:=m/T$, then
\begin{eqnarray}
\label{aa} {\cal X} \propto (-t)^\beta\,,\; && t\to 0^-\,,\;h=0\,,\\
\label{ab} {\cal X} \propto h^{1/\delta}\,,\;&& h\to 0^+\,,\;t=0\,.
\end{eqnarray}
Calculating the critical exponents is an important contemporary goal because
of the widely conjectured notion of {\it universality}, which states that
their values depend only on the symmetries of the theory, the dimension of
space, and whether or not the interaction is short-range.  In this case many
theories could be grouped into a single universality class labelled by
$(\beta,\delta)$, and chiral symmetry restoration in QCD might be describable
by a much simpler model.

The success of the nonlinear $\sigma$-model in describing long-wavelength
pion dynamics underlies a conjecture~\cite{pisarski} that the restoration of
chiral symmetry at finite-$T$ in 2-flavour QCD is characterised by a free
energy whose interaction is just that of the 3-dimensional, $N=4$ Heisenberg
magnet ($O(4)$ model).  This theory has been studied extensively and its
critical exponents are~\cite{neqfour}: $\beta^H= 0.38 \pm 0.01$, $\delta^H =
4.82 \pm 0.05$.

The conjecture that 2-flavour QCD is in the $O(4)$ universality class was
reconsidered in Ref.~\cite{kogut}, where it is argued that the compositeness
of QCD's mesons affects the nature of the phase transition.  The Gross-Neveu
model, defined by the Euclidean action
\begin{equation}
\int d^{d+1}x\,\left[\bar\psi (i\gamma\cdot \partial + m + g \sigma)\psi
        - \case{1}{2} \sigma^2\right]\,,
\end{equation}
is presented as a counter-example.  Arguments kindred to those of
Ref.~\cite{pisarski} suggest that the chiral transition in this model should
be described by a $d$-dimensional Ising model because of the discrete $\sigma
\to -\sigma$ symmetry, which is manifest after integrating out the fermions.

At $T=0$ the Gross-Neveu model exhibits chiral symmetry restoration as the
coupling $g\to g_c^+$ and at leading order in a $1/N_F$-expansion, where
$N_F$ is the number of fermions, the critical exponents are easily calculated
by exploiting the divergence that appears as the gap vanishes in the gap
equation: $\beta_{T=0}^{\rm GN}=1/(d-1)$, $\delta_{T=0}^{\rm GN} = d$.  The
same analysis applies at finite-$T$ but in this case the absence of fermion
Matsubara-frequency zero-modes ensures that the integrand in the gap equation
is finite as the gap vanishes, so that: $\beta_{T\neq 0}^{\rm GN}= \beta^{\rm
MF}= (1/2)$, $\delta_{T \neq 0}^{\rm GN} = \delta^{\rm MF} = 3$.  These
exponents are apparently unchanged by $1/N_F$-corrections~\cite{kogut} in
which case the finite temperature transition is characterised by mean field
exponents, independent of the dimension, {\it not} Ising model exponents.  If
this is correct then a description of the transition in terms of elementary
meson fields is inadequate in the vicinity of the second-order phase
transition~\cite{kogut}.

However, in a subsequent study~\cite{stephanov} of a Yukawa model, which is
equivalent to the Gross-Neveu model in the large-$N_F$ limit, it was argued
that the chiral transition {\it is} described by the expected Ising model
exponents but only on a domain of width$\,\sim\,1/N_F$ in the vicinity of the
critical-temperature and/or -external-field, with mean field exponents
describing the evolution of the order parameters outside that domain.  In
this case $1/N_F$-corrections do appear to be important.

These simple examples illustrate why the universality class containing QCD
remains uncertain and why it will be difficult to determine if, as is
plausible, the QCD chiral transition exhibits similar complexities; e.g.,
mean field exponents which may, or may not, evolve into $O(4)$ exponents very
near the critical-coupling and/or chiral limit.  We explore this question by
searching for a systematic trend in a comparison of $(\beta,\delta)$
calculated in a range of classes of Dyson-Schwinger equation models of QCD.

Calculating the critical exponents directly from Eqs.$\,$(\ref{aa}) and
(\ref{ab}) is often difficult because of numerical noise near the critical
temperature.  Another method is to consider~\cite{arnea} the chiral and
thermal susceptibilities:
\begin{equation}
\label{defchih}
\chi_h(t,h) := 
\left.\frac{\partial\, {\cal X}(t,h)}
        {\!\!\!\!\!\!\partial h}\right|_{t}\,,\;
\chi_t(t,h) :=
\left.\frac{\partial\, {\cal X}(t,h)}
        {\!\!\!\!\!\!\partial t}\right|_{h}\,.
\end{equation}
At each $h$, $\chi_h(t,h)$ and $\chi_t(t,h)$ are smooth functions of $t$ with
maxima at the pseudocritical points $t_{\rm pc}^h$ and $t_{\rm pc}^t$, where
\begin{equation}
\label{pchpct}
t_{\rm pc}^h \propto\,  h^{1/(\beta \delta)}\,
\propto\,  t_{\rm pc}^t\,.
\end{equation}
Since $\beta\delta >0$, the pseudocritical points approach the critical
point, $t=0$, as $h\to 0^+$ and
\begin{eqnarray}
\label{deltaslope}
\chi_h^{\rm pc} & := & \chi_h(t_{\rm pc}^h,h) \propto h^{-z_h}\,,\;
        z_h:= 1 - \case{1}{\delta} \,,\\
\label{betaslope}
\chi_t^{\rm pc} & := & \chi_t(t_{\rm pc}^t,h) 
        \propto h^{-z_t}\,,
        \;z_t:= \case{1}{\beta\delta}\,(1-\beta)\,.
\end{eqnarray}
Therefore, by locating the pseudocritical points and plotting the peak-height
of the susceptibilities as a function of $h$ one can determine $T_c$, $\beta$
and $\delta$.

This method was employed~\cite{kl94} in numerical simulations of finite-$T$
lattice-QCD on an $8^3\times 4$-lattice with three values of the
current-quark mass: $a \,m^{\rm lat}_q = 0.075,0.0375,0.02$, which for $1/a
\sim 4 \,T_c$, $T_c\approx 150\,$MeV, correspond to $m^{\rm lat}_q = 45, 23,
12\,$MeV, and yielded: $\beta^{\rm lat}= 0.30 \pm 0.08$, $\delta^{\rm lat} =
4.3 \pm 0.5$.  The values are consistent with a straightforward realisation
of the $O(4)$ hypothesis, however, the errors are large.  The situation
deteriorates further when more recent simulations on larger lattices and with
smaller quark masses are taken into account~\cite{el98}.  A value of
$1/\delta^{\rm lat} = 0$ is consistent with the new data, and this can
be interpreted as a signal of a {\it first order} transition.  Our
calculations indicate that this unexpected result may be an artefact, due in
part to the large values of the quark mass that contemporary lattice-QCD
simulations are restricted to.

Herein we analyse $\chi_h(t,h)$ and $\chi_t(t,h)$ in a class of confining DSE
models that underlies many successful phenomenological applications at both
zero and finite-$(T,\mu)$~\cite{pctcdr}.  The foundation of our study, as for
many others~\cite{reinhard,basti,bastiscm,prl,greg}, is the renormalised
quark DSE
\begin{eqnarray}
S^{-1}(p_{\omega_k}) & :=& i\vec{\gamma}\cdot \vec{p} \,A(p_{\omega_k})
+ i\gamma_4\,\omega_k \,C(p_{\omega_k} ) + B(p_{\omega_k} )\\ 
&= &Z_2^A
\,i\vec{\gamma}\cdot \vec{p} + Z_2 \, (i\gamma_4\,\omega_k + m_{\rm bm})\, +
\Sigma^\prime(p_{\omega_k} ).
\label{qDSE} 
\end{eqnarray}
Here $(p_{\omega_k}):= (\vec{p},\omega_k)$ with $\omega_k= (2 k + 1)\,\pi T$
the fermion Matsubara frequency, and $m_{\rm bm}$ is the Lagrangian
current-quark bare mass.  The regularised self energy is
\begin{eqnarray}
\label{sigmap}
\Sigma^\prime(p_{\omega_k}) &=& i\vec{\gamma}\cdot
\vec{p}\,\Sigma_A^\prime(p_{\omega_k} ) 
+ i\gamma_4\,\omega_k\,\Sigma_C^\prime(p_{\omega_k} ) + 
\Sigma_B^\prime(p_{\omega_k})\,, \\
\Sigma_{\cal F}^\prime(p_{\omega_k}) & =& \int_{l,q}^{\bar\Lambda}\,
\case{4}{3}\,g^2\,D_{\mu\nu}(\vec{p}-\vec{q},\omega_k-\omega_l)
\,\case{1}{4}{\rm tr}\left[{\cal P}_{\cal F} \gamma_\mu
S(q_{\omega_l})\Gamma_\nu(q_{\omega_l};p_{\omega_k})\right]\,,
\label{regself}
\end{eqnarray}
where: ${\cal F}=A,B,C$; $A,B,C$ are functions only of $|\vec{p}|^2$ and
$\omega_k^2$; ${\cal P}_A:= -(Z_1^A/|\vec{p}|^2)i\vec{\gamma}\cdot \vec{p}$,
${\cal P}_B:= Z_1 $, ${\cal P}_C:= -(Z_1/\omega_k)i\gamma_4$; and
$\int_{l,q}^{\bar\Lambda}:=\, T
\,\sum_{l=-\infty}^\infty\,\int^{\bar\Lambda}d^3q/(2\pi)^3$, with
$\int^{\bar\Lambda}$ a mnemonic to represent a translationally invariant
regularisation of the integral and $\bar\Lambda$ the regularisation
mass-scale.  In renormalising the quark DSE we require
\begin{equation}
\label{subren}
\left.S^{-1}(p_{\omega_0})\right|_{|\vec{p}|^2+\omega_0^2=\zeta^2} = 
        i\vec{\gamma}\cdot \vec{p} + i\gamma_4\,\omega_0 + m_R\;,
\end{equation}
and hence the renormalised self energies are
\begin{equation}
\label{renself}
\begin{array}{rcl}
{\cal F}(p_{\omega_k};\zeta) & = & 
\xi_{\cal F} + \Sigma_{\cal F}^\prime(p_{\omega_k};{\bar\Lambda})
    - \Sigma_{\cal F}^\prime(\zeta^-_{\omega_0};{\bar\Lambda})\,,
\end{array}
\end{equation}
$(\zeta^-_{\omega_0})^2 := \zeta^2 - \omega_0^2$, ${\cal F}=A,B,C$, $\xi_A =
1 = \xi_C$, and $\xi_B=m_R(\zeta)$.

Equations~(\ref{qDSE})-(\ref{renself}) define the exact QCD {\it gap
equation}.  It has obvious qualitative similarities to the Gross-Neveu gap
equation, in particular the explicit dependence on $\omega_k$.

In Eq.~(\ref{regself}), $\Gamma_\nu(q_{\omega_l};p_{\omega_k})$ is the
renormalised dressed-quark-gluon vertex whose structure at $T=0$ has been
much considered~\cite{ayse97}.  It is a connected, irreducible three-point
function that should not exhibit light-cone singularities in covariant
gauges; i.e., it should be regular at $(\vec{p}-\vec{q})^2 +
(\omega_k-\omega_l)^2=0$~\cite{hawes}.  A number of {\it Ans\"atze} with this
property have been proposed and employed, and it has become clear that the
judicious use of the rainbow truncation
($\Gamma_\nu(q_{\omega_l};p_{\omega_k}) = \gamma_\nu$) in Landau gauge
provides phenomenologically reliable results~\cite{mr97}.  We employ it
herein, in which case a mutually consistent constraint is $Z_1 = Z_2$ and
$Z_1^A = Z_2^A$.  The rainbow truncation is the leading term in a
$1/N_c$-expansion of $\Gamma_\nu(q_{\omega_l};p_{\omega_k})$.

$D_{\mu\nu}(p_{\Omega_k})$ in Eq.~(\ref{regself}) is the renormalised
dressed-gluon propagator, which has the form
\begin{equation}
g^2 D_{\mu\nu}(p_{\Omega_k}) = 
P_{\mu\nu}^L(p_{\Omega_k} ) \Delta_F(p_{\Omega_k} ) + 
P_{\mu\nu}^T(p_{\Omega_k}) \Delta_G(p_{\Omega_k}  ) \,,
\end{equation}
\begin{eqnarray}
P_{\mu\nu}^T(p_{\Omega_k}) & := &\left\{
\begin{array}{ll}
0, &  \mu\;{\rm and/or} \;\nu = 4,\\
\displaystyle
\delta_{ij} - \frac{p_i p_j}{p^2}, &  \mu,\nu=i,j\,=1,2,3\;,
\end{array}\right.
\end{eqnarray}
with $P_{\mu\nu}^T(p_{\Omega_k}) + P_{\mu\nu}^L(p_{\Omega_k}) =
\delta_{\mu\nu}- p_\mu p_\nu/{\sum_{\alpha=1}^4 \,p_\alpha p_\alpha}$;
$\mu,\nu= 1,\ldots, 4$.  ($\Omega_k:= 2\pi k T$ is the boson Matsubara
frequency.)  The primary class of DSE models we consider is that in which the
long-range part of the interaction, $g^2 D_{\mu\nu}(p_{\Omega_k})$, is an
integrable infrared singularity~\cite{mn83}.  We write
\begin{eqnarray}
\label{uvpropf}
\Delta_F(p_{\Omega_k}) & = & {\cal D}(p_{\Omega_k};m_g)\,,\;
\Delta_G(p_{\Omega_k})  =  {\cal D}(p_{\Omega_k};0)\,,\\
\label{delta}
 {\cal D}(p_{\Omega_k};m_g) & := & 
        2\pi^2 D\,\case{2\pi}{T}\delta_{0\,k} \,\delta^3(\vec{p}) 
        + {\cal D}_{\rm M}(p_{\Omega_k};m_g)\,,
\end{eqnarray}
where $D$ is a mass-scale parameter and ${\cal D}_{\rm M}(p_{\Omega_k};m_g)$
may be large in the vicinity of $p_{\Omega_k}^2=0$ but must be finite.  This
type of infrared singularity is motivated by $T=0$ studies of the gluon DSE.
Studies in axial gauge~\cite{atkinson}, where ghost contributions are absent,
and in Landau gauge~\cite{pennington}, when their contributions are small,
indicate that $D_{\mu\nu}(k)$ is significantly enhanced in the vicinity of
$k^2 = 0$ relative to a free gauge-boson propagator.  In the neighbourhood of
$k^2=0$ the solution is a regularisation of $1/k^4$ as a distribution, and
the enhancement persists to $k^2 \sim 1\,$GeV$^2$ where a perturbative
analysis becomes quantitatively reliable.

The model of Refs.~\cite{bastiscm} is obtained with $D:= \eta^2/2$, ${\cal
D}_{\rm M:=A}(p_{\Omega_k};m_g) \equiv 0$, and the mass-scale
$\eta=1.06\,$GeV fixed~\cite{mn83} by fitting $\pi$- and $\rho$-meson masses
at $T=0$.  It is an ultraviolet finite model and hence the renormalisation
point and cutoff can be removed simultaneously.  A current-quark mass of
$m=12\,$MeV yields $m_\pi=140\,$MeV.  As an infrared-dominant model, it
represents the behaviour of $g^2 D_{\mu\nu}(p_{\Omega_k})$ poorly away from
$p_{\Omega_k}^2\approx 0$.  However, the artefacts this introduces are easily
identified and the model exhibits many features in common with more
sophisticated {\it Ans\"atze}.

The model of Ref.~\cite{prl} is obtained with $D:= (8/9) \, m_t^2 $ and 
\begin{equation}
\label{modelfr}
{\cal D}_{\rm M:=B}(p_{\Omega_k};m_g) = \case{16}{9}\pi^2\,
\frac{1-{\rm e}^{- s_{\Omega_k} /(4m_t^2)}}
        {s_{\Omega_k} }\,,
\end{equation}
$s_{\Omega_k}:= p_{\Omega_k}^2+ m_g^2$, where $m_g^2= (8/3)\, \pi^2 T^2$.
The mass-scale $m_t=0.69\,{\rm GeV}=1/0.29\,{\rm fm}$ marks the boundary
between the perturbative and nonperturbative domains, and was also
fixed~\cite{fr} by requiring a good description of $\pi$- and $\rho$-meson
properties at $T=0$.  At a renormalisation point of $\zeta=9.47\,$GeV,
$m_R=1.1\,$MeV yields $m_\pi=140\,$MeV.  This model adds a Coulomb-like
short-range interaction to that of Refs.~\cite{bastiscm}, thereby improving
its ultraviolet behaviour.  The ratio of the coefficients in the two terms of
Eq.~(\ref{delta}) is chosen so that the long-range effects associated with
$\delta_{0k} \delta^3(\vec{p})$ are completely cancelled at
short-distances~\cite{fr}.

We also consider a finite-$T$ extension of Ref.~\cite{mr97} defined by
\begin{equation}
\label{modelmr}
{\cal D}_{\rm M:=C}(p_{\Omega_k};m_g) = 
\frac{4\pi^2}{\omega^6} D \,s_{\Omega_k} \,
        {\rm e}^{-s_{\Omega_k} /\omega^2}
+
\frac{8\pi^2\gamma_m}
{\ln
\left[\tau + \left(1+s_{\Omega_k}/\Lambda_{\rm QCD}^2\right)^2\right]}\,
\frac{1-{\rm e}^{- s_{\Omega_k} /(4m_t^2)}}
        {s_{\Omega_k} }\,,
\end{equation}
with $\tau = e^2-1$, $\gamma_m= 12/25$, $m_g^2 = (16/5)\, \pi^2 T^2$, and
$\Lambda_{\rm QCD}^{N_f=4}=234\,$MeV.  This further improves the ultraviolet
behaviour, via the inclusion of the one-loop $\ln$-suppression at
$s_{\Omega_k} \gg \Lambda_{\rm QCD}^2$, and through the Gaussian incorporates
the intermediate-range enhancement observed in zero-temperature gluon DSE
studies.  The parameters $D$, $m_t$ and $\omega$ can be fixed at $T=0$ by
requiring a good fit to a range of $\pi$- and $K$-meson properties.  Herein
we consider two phenomenologically equivalent parameter sets that differ only
in the value of $\omega$: $D=0.78\,$GeV$^2$, $m_t=0.5\,$GeV, and:
$_1\omega= 0.6\,m_t$, $_2\omega= 1.2\,m_t$.  At $T=0$ in this model the
quark mass function, $M(p^2):=B(p^2)/A(p^2)$, evolves according to the
one-loop renormalisation group formula for $p^2 > 20\,\Lambda_{\rm QCD}^2$,
and with $_1\omega$ a renormalisation point invariant current-quark mass of
$\hat m = 6.6\,$MeV yields $m_\pi=140\,$MeV while with $_2\omega$, $\hat m =
5.7\,$MeV effects that.

An often used order parameter for dynamical chiral symmetry breaking is the
quark condensate~\cite{mr97}:
\begin{equation}
\label{qbarq}
-\langle \bar q q\rangle_\zeta:= N_c\,
         \lim_{\bar\Lambda\to \infty}
        Z_4(\zeta,\bar\Lambda)\,
        \int_{l,p}^{\bar\Lambda}\,
        {\rm tr}_D\left[S_0(p_{\omega_l})\right]\,,
\end{equation}
for each massless quark flavour.  Here the subscript ``$0$'' denotes that the
dressed-quark propagator is a chiral limit solution of Eq.~(\ref{qDSE}), and
$Z_4(\zeta,\bar\Lambda)$ is the mass renormalisation constant:
$Z_4(\zeta,\bar\Lambda) \,m_R(\zeta) = Z_2(\zeta,\bar\Lambda)\, m_{\rm
bm}(\bar\Lambda)$.  There are other, equivalent order parameters and from
Eq.~(\ref{qbarq}) it follows that one such is ${\cal X} :=
B_0(\vec{p}=0,\omega_0)$.  This being a {\it bona fide} order parameter is
particularly useful and important because it means that the lowest Matsubara
frequency completely determines the character of the chiral phase transition.

The quark DSE obtained with ${\cal D}_A$ is an algebraic equation.  Its
solution is therefore easy to analyse and either directly, via
Eqs.~(\ref{aa}) and (\ref{ab}), or using the susceptibilities, it is
straightforward to establish~\cite{arnea} that this model has mean field
critical exponents and to determine the critical temperature in
Table~\ref{taba}.  The exponents are {\it unchanged} (e.g., ${\cal X}(0,h)^3$
is linear on $h\in [10^{-10},10^{-6}]$) and $T_c$ reduced by $<\,$2\% ($T_c =
\eta/(2\pi)= 0.15915\,\eta \to 0.15646\,\eta$) upon the inclusion of the
higher-order $1/N_c$-corrections to the dressed-quark-gluon vertex discussed
explicitly in Refs.~\cite{brs96}.

The critical behaviour of the ${\cal D}_B$-model is harder to explore because
the solution of the quark DSE must be obtained numerically.  The apparently
simple interaction is actually more complicated to study than that defined
with ${\cal D}_C$ because the ultraviolet $\ln[s_{\Omega_k}]$-suppression
characteristic of asymptotically free theories is lacking.  Therefore the
renormalisation group properties are those of quenched QED, which has notable
pathologies~\cite{qed4ren}.

Chiral symmetry restoration in this model was studied in Ref.~\cite{arnea}.
Herein, however, we report results obtained with an improved numerical
procedure.  The most significant change, implemented in order to make
recovery of the $T=0$ limit easier, was to place a fixed ultraviolet cutoff,
$_4\Lambda$, on $_4q^2:= \vec{q}\,^2 + \omega_l^2$.  This requires that a
different three-momentum cutoff and grid be used for each Matsubara
frequency.  For small-$T$ this corresponds closely to an $O(4)$ invariant
cutoff procedure, and for all $T$ it provides for an accurate discrete
representation of the kernel.  In the calculations we chose $_4\Lambda\sim
1.5\, \zeta \approx 14\,$GeV, and for a given value of $T$ the number of
Matsubara modes required, $l_{\rm max}$, is determined from $\omega_{l_{\rm
max}} \approx\, _4\Lambda$.

We employ two chiral order parameters in analysing chiral symmetry
restoration
\begin{equation}
{\cal X}:= B(\vec{p}=0,\omega_0), \; 
{\cal X}_C:= \frac{B(\vec{p}=0,\omega_0)}{C(\vec{p}=0,\omega_0)}.
\end{equation}
They should be equivalent and, as we will see, the onset of that equivalence
is a good way to determine the $h$-domain on which
Eqs.~(\ref{pchpct})-(\ref{betaslope}) are valid.  Further, we have verified
numerically that for $m=0$ and $t \sim 0$: $f_\pi \propto \langle\bar q
q\rangle \propto {\cal X}(t,0)$; i.e., that these quantities are all {\it
bona fide} order parameters.  It thus follows from the pseudoscalar mass
formula~\cite{mr97}: $f_\pi^2\,m_\pi^2 = 2\,m_R(\zeta)\langle\bar q q
\rangle_\zeta^0$, that $m_\pi$ diverges at the critical
temperature\cite{prl},
\begin{equation}
m_\pi^2 \propto (-t)^{-\beta}\,,\;t\to 0^-\,.
\end{equation}

The critical temperature in Table~\ref{taba} was obtained\footnote{
In these calculations we corrected a numerical error in
Refs.\protect\cite{arnea,prl}: the Debye mass was a factor of three too large
therein. That is the origin of the 15\% increase in $T_c$, which also serves
to illustrate the small influence of the perturbative Debye mass on the
results.}
using Eqs.~(\ref{pchpct}).  Its value is insensitive to whether $t_{\rm
pc}^h$ or $t_{\rm pc}^t$ is used and also to the order parameter.
$\chi_h^{\rm pc}(h)$ and $\chi_t^{\rm pc}(h)$, obtained using both order
parameters, are depicted in Fig.~\ref{frchi} and, following
Eqs.~(\ref{deltaslope}) and (\ref{betaslope}), the critical exponents are
determined from these curves.  In comparison with Ref.~\cite{arnea} it is
obvious that there is significantly less error in the function values.  This
makes it possible to distinguish the nonzero curvature on the domain $-4.5 <
\log_{10} h < -3$, whereas in Ref.~\cite{arnea} the error in the function
values was such that the scaling relations,
Eqs.~(\ref{pchpct})-(\ref{betaslope}), appeared to be valid on this domain.
In fact, the scaling relations are not valid until
\begin{equation}
\label{mfr}
\log_{10} (h/h_u)< -7\,, 
\end{equation}
$h_u:= m_R/T_c$ is defined with the current-quark mass that gives $m_\pi
=140\,$MeV in this model.

The scaling domain is most easily identified by defining a ``local'' critical
exponent for each of the equivalent chiral order parameters:
\begin{equation}
\label{zi}
z_i:= \,-\,\frac{ \ln \chi^{\rm pc}_i - \ln \chi^{\rm pc}_{i+1}}
        {\ln h_i - \ln h_{i+1}}\,,
\end{equation}
where $(h_i,\chi^{\rm pc}_i)$ and $(h_{i+1},\chi^{\rm pc}_{i+1})$ are
adjacent data pairs.  $z_i(h)$ for the present model is depicted in
Fig.~\ref{findiff}.  $h$ lies in the scaling region when $z_i$ is independent
of the order parameter.  

Anticipating the curvature, we fit the susceptibilities using the functional
form
\begin{equation}
\log_{10}\chi^{\rm pc} = \frac{1 + a_0\, \varsigma + a_1 \,\varsigma^2}
                {a_2 + a_3 \,\varsigma}\,, 
\end{equation}
$\varsigma= \log_{10}h$, over the smallest five values of $\varsigma$, and
the critical exponent is then: $z = a_1/a_3$.  We estimate the error by
systematically eliminating one value of $\varsigma$ at a time from the fit
and observing the change in $z$.  In this way we obtain the values of $z_h$
and $z_t$ in Table~\ref{taba}.

A linear fit on $\varsigma \in [-4.5,-3]$ yields: $z_h=0.79$, $z_t=0.40$,
which may be compared with the values in Ref.~\cite{arnea}: $z_h=0.77\pm
0.02$, $z_t=0.28 \pm 0.04$.  Comparing the values of $z_h$ highlights the
error introduced by incorrectly identifying the scaling region.  The
comparison between the values of $z_t$ indicates our significant improvement
of the numerical method.

The study of chiral symmetry restoration in the model obtained with ${\cal
D}_C$ is straightforward, with the additional $\ln[s_{\Omega_k}]$-suppression
in the ultraviolet making the numerical analysis much simpler.  The critical
temperature for each parameter set is presented in Table~\ref{taba}, and on
this $\omega$-domain: $T_c(\omega) \approx 88 + 107\,\omega$.  $\chi_h^{\rm
pc}(h)$ and $\chi_t^{\rm pc}(h)$ for the smallest value of $\omega$, which
corresponds to the parameter set in Ref.~\cite{mr97}, are depicted in
Fig.~\ref{mrachi}.  The results for the larger value of $\omega$ are similar.
In these cases the scaling relations are only valid for
\begin{eqnarray}
\label{mmrammrb}
_1\omega \! : \,\log_{10} (h/h_u)< -5\,, &\; &
_2\omega \! : \,\log_{10} (h/h_u)< -6\,.
\end{eqnarray}

It is clear from Table~\ref{taba} that each of these models is mean field in
nature.  In hindsight that may appear unsurprising because the long-range
part of the interaction is identical in each case and the correlation length
diverges as $t\to 0$.  However, the models differ in detail by the manner in
which the interactions approach their long-range limits, and our numerical
demonstration of their equivalence required a very careful analysis and
extremely small values of the current-quark mass, Eqs.~(\ref{mfr}) and
(\ref{mmrammrb}).

This last observation is also likely to be true in QCD; i.e., while the
critical temperature is relatively easy to determine, very small
current-quark masses may be necessary to accurately calculate the critical
exponents from the chiral and thermal susceptibilities.  If that is the case,
the calculation of these exponents via numerical simulations of lattice-QCD
will not be feasible.  The discrepancies described in Ref.~\cite{el98} could
be a signal of this.

The class of models we have considered explicitly preserves the chiral
Ward-Takahashi identity and can describe the long-wavelength dynamics of QCD
very well~\cite{pctcdr,mr97}.  Importantly, it describes that dynamics in
terms of mesons that are quark-antiquark {\it composites}.  The
characteristic feature of this class is the behaviour of the long-range,
confining interaction.  It provides an additive, algebraic driving term in
the quark DSE that is proportional to the dressed-quark propagator, which
means that boson Matsubara-frequency zero-modes do not influence the critical
behaviour determined from the gap equation.

The class of Coulomb gauge models considered in Ref.~\cite{reinhard} also
describes mesons as composite particles and it too exhibits mean field
critical exponents.  The long-range part of the interaction in that class of
models corresponds directly to the regularised Fourier amplitude of a
linearly rising potential. Hence it is not equivalent to ours in any simple
way, except insofar as zero modes do not influence the gap equation.

We have also considered the class of models obtained with~\cite{basti}:
${\cal D}(p_{\omega_k}-q_{\omega_l};m_g)\propto g(|\vec{p}|)\,g(|\vec{q}|)$,
where $g(|\vec{p}|)$ is a non-increasing function of its argument.  This
class includes the non-confining NJL-model, which is obtained with
$g(|\vec{p}|)= \theta(1-|\vec{p}|/ _3\!\Lambda)$, where $_3\!\Lambda$ is a
cutoff parameter, and can also provide a good description of long-wavelength
pion dynamics.  The models in this class describe mesons as composites and
exhibit an explicit fermion substructure, and those we have explored have
mean field critical exponents.  The same is true of separable models with
$g_i=g_i(\omega_k^2+\vec{p}\,^2)$\cite{pctqt}.

The quark DSE is the QCD gap equation and the many equivalent chiral order
parameters are directly related to properties of its solution.  We have
observed that four classes of models exhibit the same (mean field) critical
exponents.  The classes are distinguished by the qualitatively distinct type
of interaction they employ in the gap equation but similar in using the
rainbow truncation.  Only in our simplest confining model did we consider the
effect of $1/N_c$-corrections to the quark-gluon vertex, and in that case the
critical exponents were unchanged.  Incipient in these results is the
suggestion that mean field exponents are a feature of the essential fermion
substructure in the gap equation, and not of the interaction.\footnote{
Our discussion is independent of the number of light quark flavours insofar
as that number does not suppress the interaction strength to such an extent
that chiral symmetry is not dynamically broken at $T=0$.}
If that is correct then chiral symmetry restoration at finite-$T$ in QCD is a
mean field transition.  This hypothesis is supported by chiral random matrix
models of QCD, whose formulation relies only on the symmetries of the Dirac
operator and which also yield~\cite{jackson96wettig97} mean field exponents.
It can likely only be false if $1/N_c$-corrections to the vertex are large in
the vicinity of the transition, in which case their null effect in our
simplest model will have been misleading.

\acknowledgements AH acknowledges the hospitality of the Physics Division at
Argonne National Laboratory, and PM and CDR that of the Physics Department at
the University of Rostock during visits facilitating this work, which was
supported in part by: Deutscher Akademischer Austauschdienst; the US
Department of Energy, Nuclear Physics Division, under contract number
W-31-109-ENG-38; the National Science Foundation under grant no.~INT-9603385;
and benefited from the resources of the National Energy Research Scientific
Computing Center.

%

\begin{table}
\begin{tabular}{cccccc}
            & A    & B   & C$_{_1\omega}$ & C$_{_2\omega}$ \\
$T_c (MeV) $&  169 & 174 & 120 & 152  \\
$z_h$       &  0.666 & 0.67 $\pm$ 0.01  & 0.667 $\pm $ 0.001 
                        &0.669 $\pm$ 0.005 \\
$z_t$       &  0.335 & 0.33 $\pm$ 0.02  & 0.333 $\pm $ 0.001 & 0.33 $\pm$ 0.01   
\end{tabular}
\caption{Critical temperature for chiral symmetry restoration and critical
exponents characterising the second-order transition in the four exemplary
models.  Mean field critical exponents are: $z_h= 2/3$, $z_t=1/3$.
\label{taba}}
\end{table}
\setcounter{figure}{0}
\begin{figure}[h]
\centering{\ \epsfig{figure=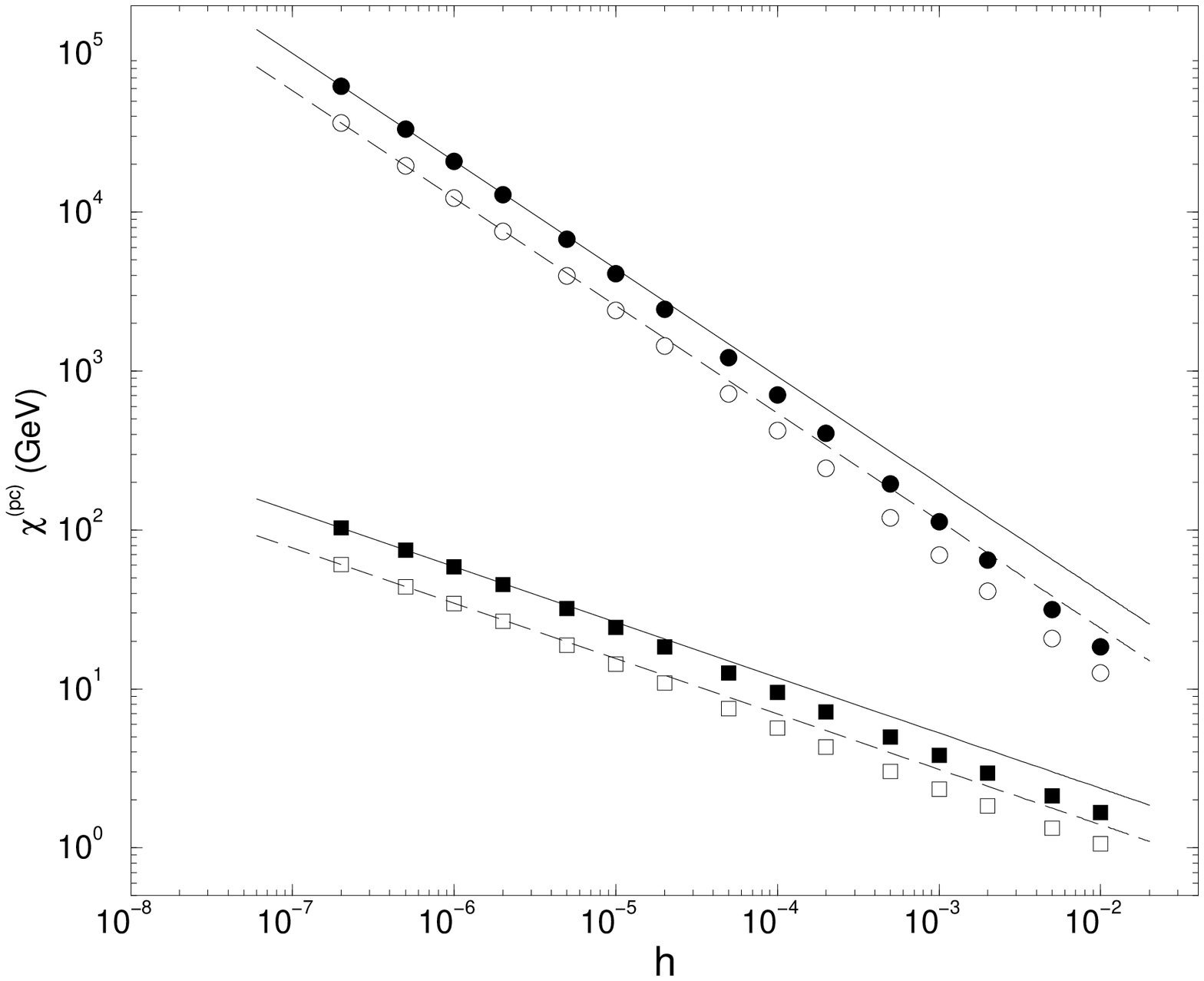,height=12.0cm}}
\caption{$\chi_h^{\rm pc}(h)$ (circles) and $\chi_t^{\rm pc}(h)$ (squares)
calculated in the model defined via ${\cal D}_B$ in
Eq.~(\protect\ref{modelfr}): filled symbols - ${\cal X}$, open symbols -
${\cal X}_C$.  The slope of the straight lines is given in
Table~\protect\ref{taba} and they are drawn through the two smallest
$h$-values.
\label{frchi}}
\end{figure}

\begin{figure}[h]
\centering{\ \epsfig{figure=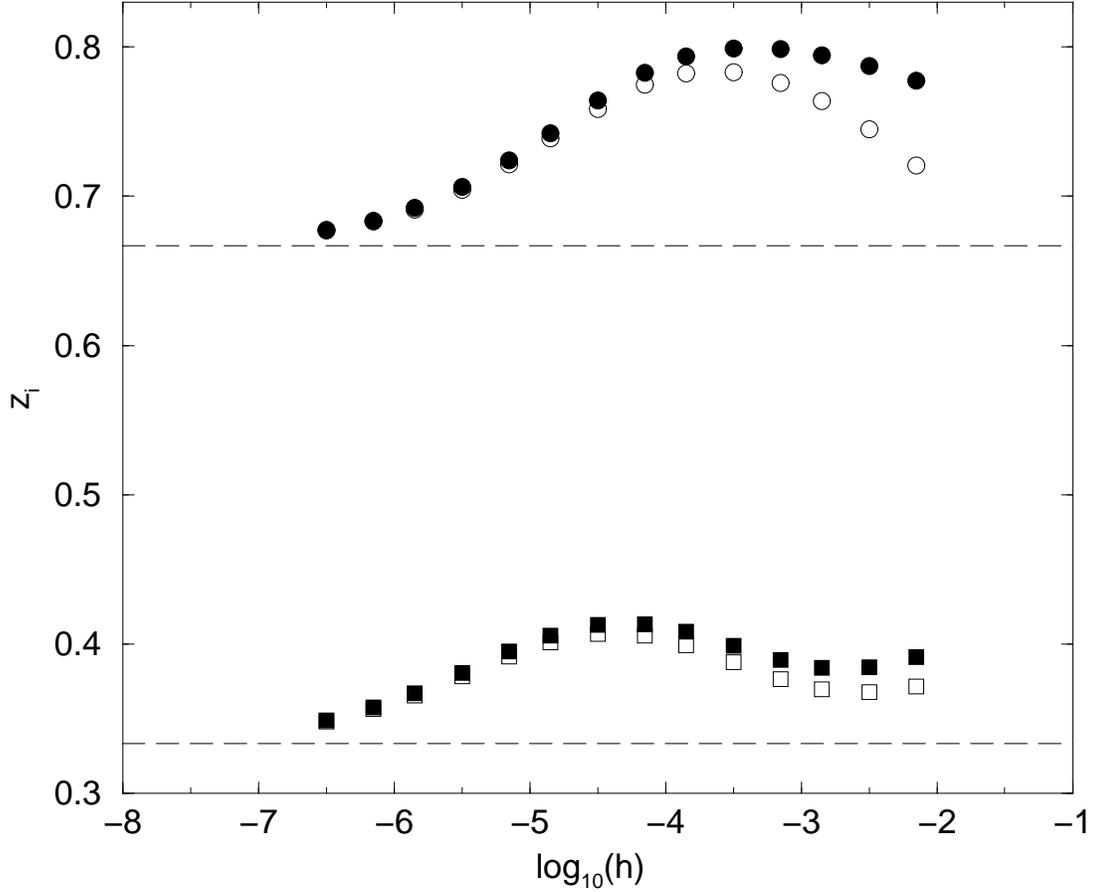,height=12.0cm}}
\caption{$z_i^h$ (circles) and $z_i^t$ (squares) from Eq.~(\protect\ref{zi})
in the model defined via ${\cal D}_B$ in Eq.~(\protect\ref{modelfr}): filled
symbols - ${\cal X}$, open symbols - ${\cal X}_C$.  The dashed lines are the
mean field values: $z_h=2/3$, $z_t=1/3$.
\label{findiff}}
\end{figure}

\begin{figure}[h]
\centering{\ \epsfig{figure=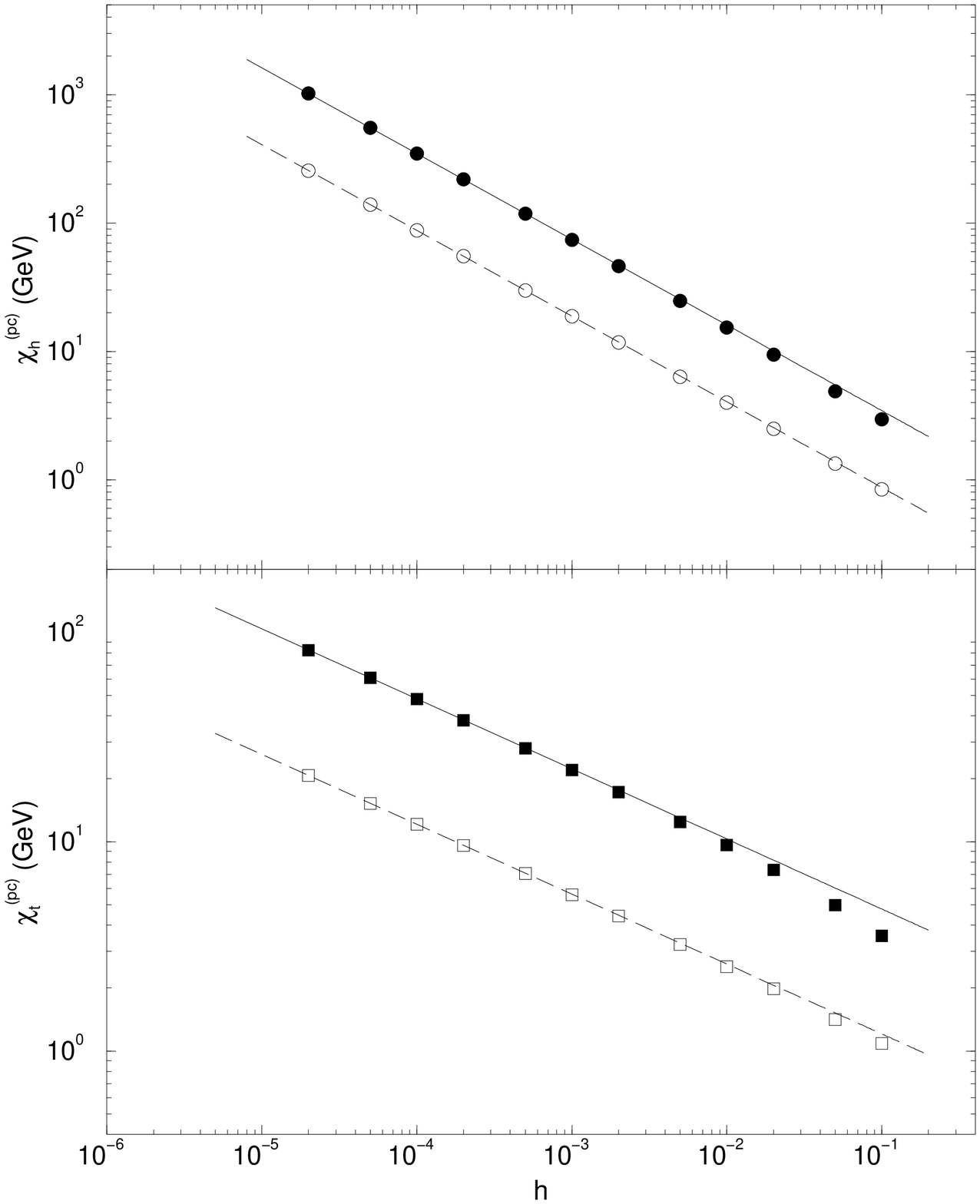,height=16.0cm}}
\caption{$\chi_h^{\rm pc}(h)$ (circles) and $\chi_t^{\rm pc}(h)$ (squares)
calculated in the model defined via ${\cal D}_C$ in
Eq.~(\protect\ref{modelmr}) with $\omega = 0.6\, m_t$: filled symbols -
${\cal X}$, open symbols - ${\cal X}_C$.  The slope of the straight lines is
given in Table~\protect\ref{taba} and they are drawn through the two smallest
$h$-values.
\label{mrachi}}
\end{figure}


\begin{thebibliography}{99}
%
\bibitem{pisarski} R. Pisarski and F. Wilczek, Phys. Rev. D {\bf 29}, 338
(1984); F. Wilczek, Int. J. Mod. Phys. A {\bf 7}, 3911 (1992). 
%
\bibitem{neqfour} G. Baker, B. Nickel and D. Meiron, Phys. Rev. B {\bf 17},
1365 (1978); and ``Compilation of 2-pt. and 4-pt. graphs for continuous spin
models'', University of Guelph report (1977), unpublished.
%
\bibitem{kogut} A. Koci\'c and J. Kogut, Phys. Rev. Lett. {\bf 74}, 3109
(1995); Nucl. Phys. B {\bf 455}, 229 (1995).
%
\bibitem{stephanov} J. B. Kogut, M. A. Stephanov and C. G. Strouthos,
``Critical region of the finite temperature chiral transition'',
hep-lat/9805023
%
\bibitem{arnea} D. Blaschke, A. H\"oll, C. D. Roberts and S. Schmidt,
Phys. Rev. C {\bf 58}, 1758 (1998).
%
\bibitem{kl94} F. Karsch and E. Laermann, Phys. Rev. D {\bf 50}, 6954 (1994).
%
\bibitem{el98} E. Laermann, Nucl. Phys. B {\bf 63} (Proc. Suppl.), 114
(1998).
%
\bibitem{pctcdr} P. C. Tandy, Prog. Part. Nucl. Phys. {\bf 39}, 117 (1997);
C. D. Roberts, ``Nonperturbative QCD with Modern Tools'', nucl-th/9807026.
%
\bibitem{reinhard} R. Alkofer, P. A. Amundsen and K. Langfeld, Z. Phys. C
{\bf 42}, 199 (1989).
%
\bibitem{basti} S. Schmidt, D. Blaschke and Yu. L. Kalinovsky, Phys. Rev. C
{\bf 50}, 435 (1994).
%
\bibitem{bastiscm} P. Maris, C. D. Roberts and S. Schmidt, Phys. Rev. C {\bf
57}, R2821 (1998); D. Blaschke, C.D. Roberts and S. Schmidt, Phys. Lett. B
{\bf 425}, 232 (1998).
%
\bibitem{prl} A. Bender, D. Blaschke, Yu. Kalinovsky and C.D. Roberts,
Phys. Rev. Lett. {\bf 77}, 3724 (1996). 
%
\bibitem{greg} A. Bender, {\it et al}., Phys. Lett. B {\bf 431}, 263 (1998)
pp. 263-269
%
\bibitem{ayse97} A. Bashir, A. Kizilersu and M.R. Pennington, Phys. Rev. D
{\bf 57}, 1242 (1998); and references therein.
%
\bibitem{hawes} F. T. Hawes, P. Maris and C. D. Roberts, ``Infrared Behaviour
of Propagators and Vertices'', nucl-th/9807056, Phys. Lett. B., in press.
%
\bibitem{mr97} P. Maris and C. D. Roberts, Phys. Rev. C {\bf 56}, 3369
(1997).
%
\bibitem{mn83} H.J. Munczek and A.M. Nemirovsky, Phys. Rev. D {\bf 28}, 3081
(1983).
%
\bibitem{atkinson} M. Baker, J. S. Ball and F. Zachariasen, Nucl. Phys. B
{\bf 186} 531 (1981); {\it ibid} 560; D.~Atkinson, P. W. Johnson,
W. J. Schoenmaker and H. A.  Slim, Nuovo Cimento A {\bf 77}, Series 11 (1983)
197.
%
\bibitem{pennington} N. Brown and M. R. Pennington, Phys. Rev. D {\bf 39},
2723 (1989); M. R. Pennington, ``Calculating hadronic properties in strong
QCD'', hep-ph/9611242.
%
\bibitem{fr} M.R. Frank and C.D. Roberts, Phys. Rev. C {\bf 53}, 390 (1996).
%
\bibitem{brs96} A. Bender, C. D. Roberts and L. v. Smekal, Phys. Lett. B {\bf
380} (1996) 7; C. D. Roberts, in {\it Quark Confinement and the Hadron
Spectrum II}, edited by N. Brambilla and G. M. Prosperi (World Scientific,
Singapore, 1997), pp. 224-230.
%
\bibitem{qed4ren} F. T. Hawes, T. Sizer and A. G. Williams, Phys. Rev. D {\bf
55}, 3866 (1997).
%
\bibitem{pctqt} P. C. Tandy, private communication.
%
\bibitem{jackson96wettig97} A. D. Jackson and J. J. M. Verbaarschot,
Phys. Rev. D {\bf 53}, 7223 (1996).  
%
\end{thebibliography}
\end{document}